\documentclass[conference]{IEEEtran}
\usepackage{amsmath,amssymb,amsfonts}
\usepackage{algorithmic}
\usepackage{graphicx}
\usepackage{textcomp}
\usepackage{xcolor}
\usepackage{subfig}
\usepackage{float}
\usepackage{xcolor}
\usepackage{soul}
\def\BibTeX{{\rm B\kern-.05em{\sc i\kern-.025em b}\kern-.08em
    T\kern-.1667em\lower.7ex\hbox{E}\kern-.125emX}}

\usepackage{fancyhdr}
\begin{document}

\title{``I think you need help! Here's why'': Understanding the Effect of Explanations on Automatic Facial Expression Recognition}

\author{\IEEEauthorblockN{1\textsuperscript{st} Sanjeev Nahulanthran}
\IEEEauthorblockA{\textit{Faculty of Information Technology} \\
\textit{Monash University}\\
Melbourne, Australia \\
sanjeev.nahulanthran@monash.edu}
\and
\IEEEauthorblockN{2\textsuperscript{nd} Mor Vered}
\IEEEauthorblockA{\textit{Faculty of Information Technology} \\
\textit{Monash University}\\
Melbourne, Australia \\
mor.vered@monash.edu}
\and
\IEEEauthorblockN{3\textsuperscript{rd} Leimin Tian}
\IEEEauthorblockA{\textit{Faculty of Engineering} \\
\textit{Monash University}\\
\textit{\& CSIRO Robotics} \\
Melbourne, Australia \\
leimin.tian@monash.edu}
\and
\IEEEauthorblockN{4\textsuperscript{th} Dana Kuli\'c}
\IEEEauthorblockA{\textit{Faculty of Engineering} \\
\textit{Monash University}\\
Melbourne, Australia \\
dana.kulic@monash.edu}
}

\maketitle
\thispagestyle{fancy}

\begin{abstract}
    Facial expression recognition (FER) has emerged as a promising approach to the development of emotion-aware intelligent systems. The performance of FER in multiple domains is continuously being improved, especially through advancements in data-driven learning approaches. However, a key challenge remains in utilizing FER in real-world contexts, namely ensuring user understanding of these systems and establishing a suitable level of user trust towards this technology. We conducted an empirical user study to investigate how explanations of FER can improve trust, understanding and performance in a human-computer interaction task that uses FER to trigger helpful hints during a navigation game. Our results showed that users provided with explanations of the FER system demonstrated improved control in using the system to their advantage, leading to a significant improvement in their understanding of the system, reduced collisions in the navigation game, as well as increased trust towards the system.
\end{abstract}


\section{Introduction}

Facial expression recognition (FER) is a key challenge in affective computing and has shown potential to improve human-computer interaction (HCI)~\cite{dornaika2009facial} 
Despite recent advancements, several challenges remain in developing accurate, reliable, and robust FER that can operate in real-world contexts. One major challenge is the lack of transparency to the end users, which is becoming increasingly important in the context of practical affective computing research and application~\cite{tian2022aac}. A lack of transparency with regards to how an integrated FER system uses the data input to make predictions or decisions could affect a user's perception and reliance towards the system~\cite{linardatos2020explainable}. A system that is not transparent could be perceived as ambiguous or not accountable. These situations should be avoided to allow a shared understanding between the user and the system~\cite{gregor1999explanations} and avoid any mismatch of goals~\cite{abdul2018trends}.

This study aims to develop and evaluate methods for improving an off-the-shelf FER model's transparency and user understanding, thus allowing users to comprehend the performance of the FER system, allowing for a seamless and effective collaboration. Although there have been previous attempts to address inconsistencies in FER models \mbox{\cite{landowska2021mining, del2022understanding}} and improve their performance from the developer's perspective, to the best of our knowledge, there has been no existing work that focuses on facilitating end-users' understanding on the predictions of FER systems. Towards this aim, we utilize eXplainable Artificial Intelligence (XAI) and investigate its potential benefits for systems employing FER from the end-user perspective of these technologies.

Specifically, we used an off-the-shelf machine learning model for FER, utilizing Support Vector Machines (SVMs)~\cite{baltrusaitis2018openface} to predict Facial Action Units (FAUs), which feeds into an expert system~\cite{pantic2000expert} to predict categorical emotions. We also use FAUs to generate explanations, which ensures consistency between the explanation and model's decision, i.e., the features used to make the model decisions (based on FAUs) are directly used as explanations. We then developed an interactive game in which the FER system is used to trigger helpful hints to the user and compared performance, user understanding, number of hints triggered as well as accepted, and trust between cohorts who were provided with explanations about the FER system and those who were not. In the study (N=20), we compared providing explanations to the emotion-aware hint system with a control condition without explanations through a between-subject study. We then conducted human-grounded evaluation methods~\cite{doshi2017towards, hoffman2018metrics} using a combination of quantitative in-game metrics and surveys as well as qualitative interviews to determine the effect of explanations on FER towards overall system understanding and transparency.

Our study addresses the research question:
\textit{How do explanations on a FER model's prediction affect a user's \textbf{performance} in HCI, as well as their \textbf{understanding and trust} towards a HCI system's assistance?}
Based on \mbox{\cite{wobbrock2012seven}}, our work has three main contributions:
\begin{itemize}
    \item \textbf{Artifact}: We develop a framework system which can be used to test different XAI methods on HCI tasks. The tool is modular and can be used with FER systems that output the 6-categorical emotion predictions on a variety of user developed tasks. This system is open-source and shall be provided for future FER explainability research;
    \item \textbf{Empirical}: We demonstrate that users who are provided with explanations on FER have a better understanding and control of the system, leading to greater task performance. We also demonstrate that these users have a higher degree of trust towards the system;
    \item \textbf{Dataset}: We contribute a dataset containing fine-grain interaction data and participant in-game metric, survey results and interview responses for both explanation and control conditions. We also provide open-source code of the game environment for replication.\footnote{Author contact for links to the dataset and code: sanjeev.nahulanthran@monash.edu }
\end{itemize}

\section{Related Work}
We briefly review the various benefits and limitations of FER systems for HCI. We then review XAI and discuss its potential benefits in the context of FER and HCI.

\subsection{FER Systems}

FER is a widely adopted, non-contact method for emotion recognition and has garnered growing research interest in the past few years~\cite{zeng2007survey, fasel2003automatic, castellano2023automatic}. FER has been shown to be naturalistic,  unobtrusive, economical and easy to deploy and maintain using commercially available sensing systems~\cite{kolli2011non, imani2019survey, kulke2020comparison, mone2015sensing}.
One commonly adopted FER method is the Facial Action Coding System (FACS)~\cite{ekman1978facial}. This method encodes human facial expressions as facial muscle activations associated with a person's emotional state. Each FAU represents a visible facial muscle movement. We used this method as it provides a formal representation of facial expressions through muscle activations which are easy to explain and grounded in anatomy.

Literature in FER has focused on improving classification accuracy and rarely investigates the perception, reliance and trust of users~\cite{devillers2021human}
As the field continues to advance rapidly, it is crucial to provide a deeper understanding of machine learning models so that end users can have a greater sense of autonomy when using these systems~\cite{abdul2018trends}. Aside from enhancing the system's intelligence and social acceptance, empowering users to make informed decisions and retain control is equally vital. 

\subsection{XAI}

Explanations are crucial for imparting knowledge and sharing experience among people~\cite{keil2006explanation}. It can be defined in multiple ways such as \textit{an assignment of causal responsibility}~\cite{kahneman2011thinking} or simply as \textit{an answer to a ``why-question''}~\cite{
graziani2023global}. In the field of artificial intelligence (AI), explainability (also sometimes equated with interpretability)~\cite{doshi2017towards} helps to create a shared understanding~\cite{miller2019explanation} between end users and the AI system that is being used.

The two main approaches to explanation generation in XAI are \textit{intrinsic} explanations and \textit{post-hoc} explanations. Intrinsic explanations can be generated for white-box models, systems that inherently have some amount of interpretability ``built-in'' that accompanies the model's output~\cite{minh2022explainable}, whereas post-hoc methods are techniques applied ``after'' a model's prediction is made~\cite{doshi2017towards}. In our research, we focus on intrinsic explanations for any FER model that utilizes FAUs as inputs to the expression recognition system. Specifically, we use SVMs~\cite{cortes1995support} to perform FAU prediction that feeds into an expert system~\cite{pantic2000expert} for emotion prediction. We then generate  human-readable explanations for the predicted emotions based on the activated FAUs to improve system interpretability (seen in Sec.~\ref{explanation-mechanism}).

Explanation can also vary depending on the type of insight provided to users. These are, namely, \textit{Global} and \textit{Local} explanations~\cite{doshi2017towards, miller2019explanation}
. Global explanations provide the user with general knowledge, such as the features a model uses to make a decision. Local explanations on the other hand provide knowledge about the specific prediction that was made by a model. As there has been no prior literature on which explanation type would be beneficial to end-users in the context of FER, in this research, we use both global and local explanations to provide a complete understanding of our system to end-users. More details on the global and local explanation design is provided in Sec.~\ref{explanation-mechanism} as it is embedded in the testing environment.

Miller  \cite{miller2019explanation} provides insights on what constitutes a good explanation, specifically that ``\textit{probabilities and statistics don't matter}'' and that ``\textit{explanations are social}''. We incorporate these insights by showing the user which FAUs lead to a prediction accompanied by natural language explanations. This allows the explanation to take the form of a conversation or interaction with the user.

Past studies of XAI found explanations to improve users' understanding of AI systems~\cite{buccinca2020proxy, cheng2019explaining}. However, empirical results regarding its impact on users' subjective experience such as trust~\cite{
zhang2020effect,vered2023effects} and acceptance~\cite{forster2020fostering
} are varied and particularly rare when it comes to the domain of FER. 

\subsection{XAI in the context of FER}

In the context of explanations for expression recognition, very few research publications have been produced. One example of how explanations of facial expressions can be beneficial is shown in~\cite{ellingsen2022patient}, who used patient-clinician simultaneous fMRI scans synchronised with facial expression recordings in order to correlate active brain regions with FAUs associated with pain stimulus. The authors used SHAP~\cite{lundberg2017unified}, to determine the features most highly associated to pain and correlated them to studies that identified brain regions most associated with pain. While this work is relevant in understanding how pain is related to involuntary facial expressions, it may not generalize to other interaction scenarios. However, this work demonstrates that explanations can be utilised to extract context-specific knowledge necessary to better understand end-users.

Most relevant to our work are~\cite{rathod2022kids} and~\cite{del2022understanding}, in which the authors themselves received explanations (generated through Grad-CAM, SoftGrad, LIME and CEM methods) on emotions identified by CNN models and proceed to qualitatively analyse the explanations. The explanations highlight the features (superpixels) which contribute most to the predictions being made. The authors distinguish between the different XAI methods through qualitative comparison obtained by the authors themselves. However, how helpful these explanations are to an average, non-expert user remains unclear. Crucially, these studies do not empirically evaluate the perceptions of non-expert users and the usability of the explanations with regards to the emotions predicted and the underlying mechanisms of the system. Moreover, these studies do not evaluate the performance of FER models in a real world setting, nor did they measure perception and trust of the users towards the FER when exposed to explanations~\cite{tian2022aac, miller2022we}. Although there has been work in comparing some XAI techniques in the literature~\cite{del2022understanding} with regards to FER, utilizing FAUs to explain facial expression in an empirical setting has not been carried out to the best of our knowledge and their current benefits or usability are unknown.

Various evaluation methods have been proposed to evaluate the effectiveness of XAI systems from the end users' perspective~\cite{hoffman2018metrics}. These evaluations have not been conducted in the context of explanations for FER, which is the focus in this work. By moving experiments from a human-grounded approach to an application-grounded approach ~\cite{doshi2017towards}, we expect effective XAI systems can be developed to better suit the context of emotion recognition. In this work, we leverage these evaluation methods and construct an empirical user study to  evaluate the need and effects of explanations on FER systems, thereby providing a foundation for future research on explaining FER systems to end-users.

\section{Hypotheses}

The study was designed to determine the effect of explanations on user interaction with a FER System. We hypothesised that the users provided with explanations would have better understanding and control of the system, leading to higher hint acceptance and compliance. Through this greater control and performance, we expect that users will trust the system more when explanations are provided. Therefore, we investigated the following hypotheses:

\begin{enumerate}
\item     H1: A user will have better task performance when provided with explanations

\item     H2: A user will report and demonstrate better understanding of the system when provided with explanations

\item     H3: A user will report higher trust towards the system when provided with explanations
\end{enumerate}

\section{System Design}

\subsection{FER System and Hint Trigger Mechanism}\label{fer-system-hint-mechanism}

\begin{figure}[h]
    \centering
    \includegraphics[width=0.8\linewidth]{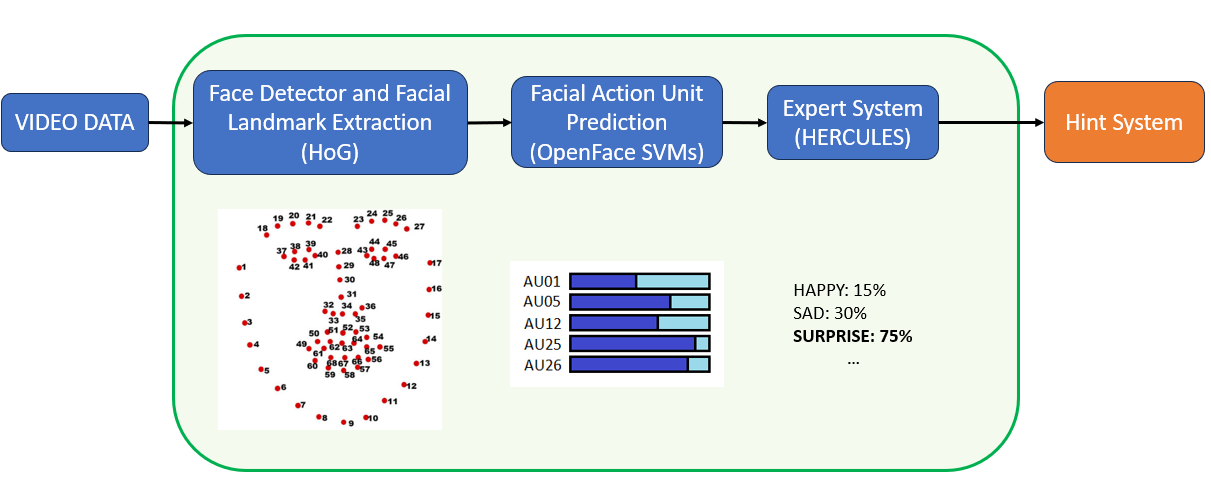}
    \caption{The Facial Expression Recognition System triggers hints based on a user's facial expressions.}
    \label{fer-detection-system-fig}
\end{figure}

We implemented a FER system that streams video data and detects faces using the face predictor implemented with the dlib library\mbox{~\cite{dlib09}}. We then utilize OpenFace's SVM library\mbox{~\cite{baltrusaitis2018openface}} to predict FAU activations from the detected facial region. FAU activations are fed into the HERCULES production rules expert system\mbox{~\cite{pantic2000expert}}. This expert system computes the confidence of the facial expression being associated with each of the ``Big-6'' emotion categories: happy, sad, angry, surprised, disgusted, fearful. If the confidence value of all emotions is below 50\%, the system classifies that facial expression as \textit{neutral}. Otherwise, the emotion with the highest confidence value is selected as the classification output. The system has an accuracy of 86.3\% when tested against annotations by 5 certified FACS coders \mbox{\cite{pantic2000expert}} and 89\% on the benchmark Cohn-Kanade dataset \mbox{\cite{fasel2003automatic}}. The output of the emotion prediction is used by the Hint System as seen in Fig.~\ref{fer-detection-system-fig}.

The Hint System stores the emotions inferred within the last 2 seconds and uses majority voting to determine when to trigger a hint for the user. If the majority of the emotions inferred within the 2 second window was negative emotions (anger, fear, sadness, disgust) or surprise, a hint is triggered.

The Hint System was tested in a separate pilot study with 20 participants who did not participate in the main study. It was conducted as a within-subject experiment to measure the usability of the hint mechanism. The pilot study tested the use of a manual hint button compared with the proposed automatic hint mechanism triggered by facial expressions. Each participant interacted with 3 versions of the system in random order – the autohint system using a 2-second timing window (decided based on previous research \cite{mcquillin2022learning, cernea2013study, panwar2018providing}), a manually triggered hint system  and a system without any hints. A pairwise Wilcoxon signed-rank sum test was used (non-normally distributed data) to compare the means of the game score, number of collisions and hints triggered between each version. The results of the test indicated no significant difference (p $>$ 0.1) between the manual and autohint system. This suggests that the autohint system has similar usability to the manual hint system and therefore we adopted the autohint system in the main study.

\subsection{Explanation Mechanism}\label{explanation-mechanism}

\subsubsection{Global explanations}

As part of the global explanation, participants were shown a video at the start of the session, which provided examples of the emotions that trigger hints, as well as FAUs that contribute to the emotion prediction. This video was shown only once prior to the start of the game and was 5 minutes in length. The purpose of global explanations was to provide a high level understanding of what emotions help to trigger the hint function, how long the emotion should be shown for effective hint triggering and which FAUs were important for each emotion. This explanation aimed to answer the question \textit{'How does the model generally work?'} corresponding to the definition of global explanations as defined in\mbox{~\cite{hoffman2018explaining}}.

\subsubsection{Local explanations}

In the case of local explanations, participants had the opportunity to review why a certain emotion prediction was made during the Explorer Game. Participants were able to view these local explanations at any time they wished (with or without the hint system being triggered) through the use of buttons.

\begin{figure}[h]
    \includegraphics[width=\linewidth]{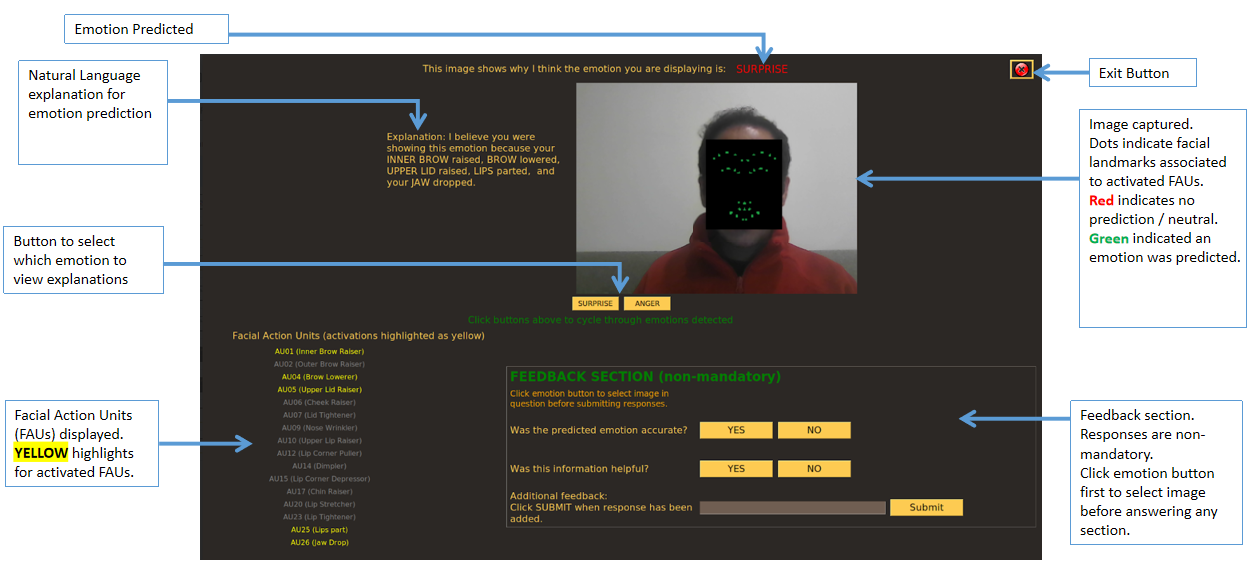}
    \caption{Local Explainer User Interface. Users can review the predicted emotion category (top of the window), the corresponding frames with facial landmark annotations (top panel, face anonymised), and facial action unit activation information (left panel).}
    \label{explainerinterface-fig}
\end{figure}

The purpose of local explanations was to help a user understand why a particular decision is made in a given situation \cite{hoffman2018explaining}. In this instance, the facial action units (FAUs) used to determine the emotional expression were also provided as  explanations to the predicted emotion. Example FAUs include: `Lips parted', `Eyebrow raised', `Nose wrinkled'. The FAUs were presented to the user both visually and textually as shown in  Fig.\mbox{~\ref{explainerinterface-fig}}.  We elected to show explanations on-demand, as opposed to all the time, to avoid overloading the user with too much information~\cite{vered2020demand}.

\section{Methodology}

\begin{figure}[h]
    \centering
    \includegraphics[width=1\linewidth]{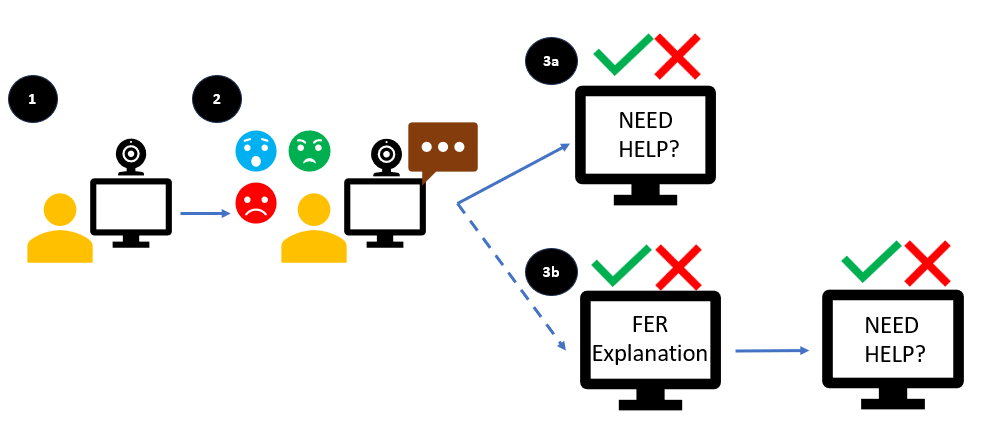}
    \caption{ Framework Overview. 1) The user interacts with a computer, while their face is visible to a camera. 2) The system provides hints in the game (pop-up) based on the user's facial expressions. 3a) Users from the control cohort are asked if they would like a hint before being provided with one. 3b) Users from the explanation cohort view the same but are also given the option to view explanations on the FER system's output. Users from both cohorts have the opportunity to accept or reject the hints triggered by their facial expression.}
    \label{general-study-flow-fig}
\end{figure}

We designed a between-subject study to investigate the effect of explanations on performance, trust and system utilization of a FER system in a navigation-based game. The system triggered helpful hints in the game based on recognized expressions. In addition to detailed objective metrics collected in-game, the experiment was followed by surveys to measure perceived trust and system usability. An overview of the system's interaction with the user is shown in Fig.~\ref{general-study-flow-fig}.

\subsection{Game Play}

\begin{figure}[tb]%
\centering
\subfloat[\label{game-interface-fig}Explorer Game Interface.]{{\includegraphics[width=0.41\linewidth]
{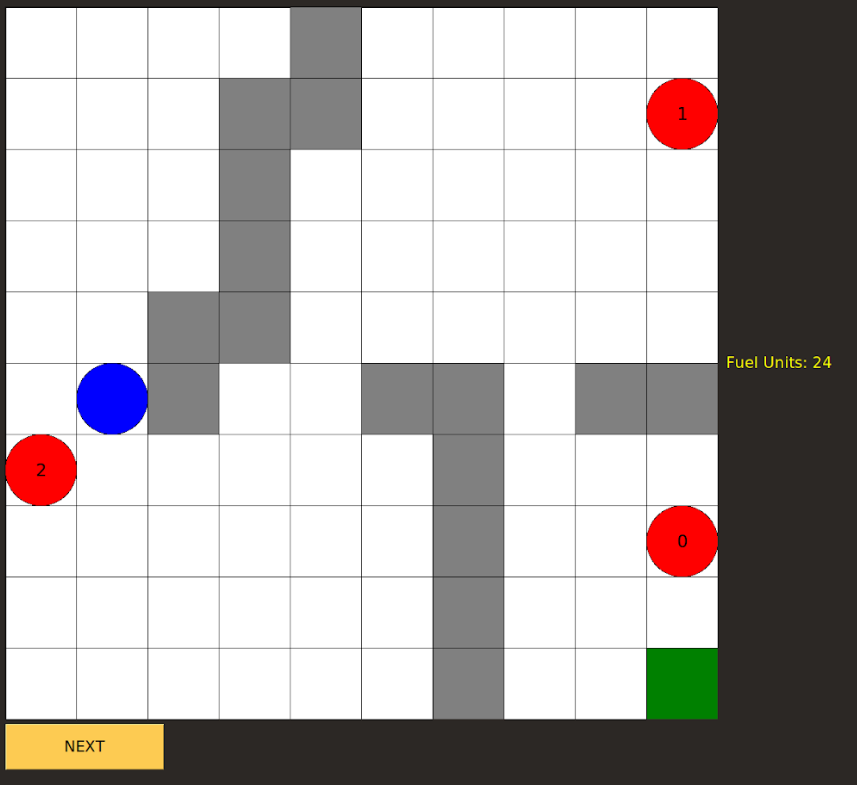} }}%
~\quad
\subfloat[\label{hintacceptcomply-fig}Top: \textit{Hint Acceptance PopUp}. \newline Bottom: \textit{Hint Compliance PopUp}]{{\includegraphics[width=0.45\linewidth]
{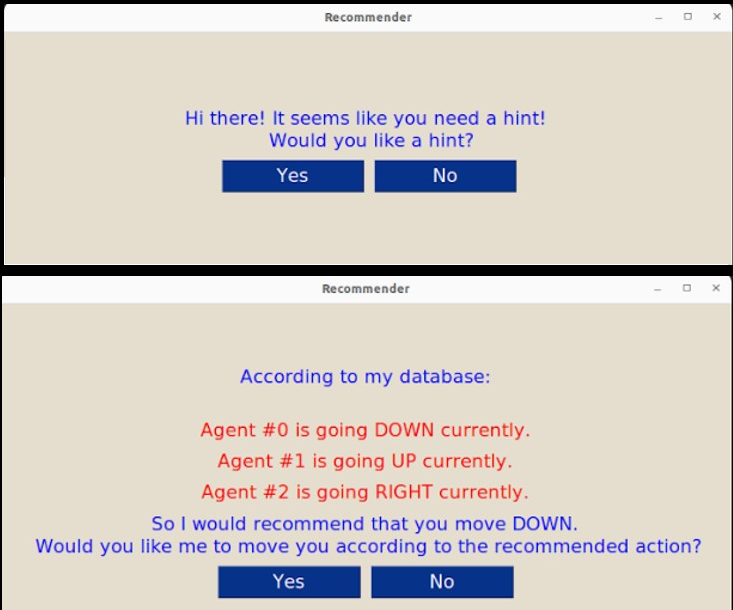} }}%
\caption{Explorer Game Interface (a) and Hint Trigger Mechanism Interface (b).}%
\label{HintTrigger}%
\end{figure}

The task chosen for this study involves a multi-agent navigation game referred to as the \textbf{Explorer Game} (see Fig.~\ref{game-interface-fig}). The participant (blue dot) navigates in a grid-world with obstacles (grey boxes) and enemy agents (red dots). The aim of the game is to navigate to the end zone (green box) while making the least number of moves possible and also avoiding collision with enemy agents or obstacles. The participant begins with 30 fuel units and can move up, down, left and right (not diagonally) or skip their turn. The enemy agents' moves are random and cannot be predicted by the participant (forcing the participant to place more reliance on the hint option which will be explained in the next section).

Each move the participant makes consumes 1 fuel unit. Colliding with an obstacle or the edges of the grid-world results in a penalty of 1 fuel unit, while colliding into an enemy agent results in a penalty of 3 fuel units. Each game instance had 16 obstacle boxes and 3 enemy agents. If the participant runs out of fuel, they lose that particular game trial and continue to the next. Winning a game involves getting the participant's agent to the end zone before the fuel runs out. The amount of remaining fuel after completing all trials determines the participants' score and was used to populate a leader-board which was shared among all participants. Each participant played 9 trials each with different maps configurations.

During the game, if the participant triggers the hint function through the mechanism explained in the previous section, the system will generate a pop-up window stating that the participant may need a hint and asking if the participant would like one (as seen in Fig.~\ref{hintacceptcomply-fig} (top)). If the participant selects ``Yes'', we consider this as hint \textbf{``acceptance''}. If the participant selects ``No'', the window will close and the participant continues to play the game. For participants who ``accepted'' the hint, they are then given information on the enemy agents' next moves as well as a recommended move to make (as seen in Fig.~\ref{hintacceptcomply-fig} (bottom)). If the participant selects ``Yes'' to follow the recommended move, the system will automatically make that move for the participant. This is defined as hint \textbf{``compliance''}.

In addition to the \textbf{Explorer Game}, we designed a \textbf{Test Game} to evaluate the participants' knowledge of the FER system. This gives us an opportunity to estimate each participant's particular understanding of how the FER works, following their experience with the Explorer Game. In the \textbf{Test Game}, there are no enemy agents (red dots),  obstacles (gray squares) or hints. The agent has a pre-planned set of moves used to get to the end zone and the keyboard is disabled. The participant can only move the agent by triggering the FER system with the same facial expressions used to trigger the hint system in the Explorer Game. The objective of the Test Game is to get as far as possible within a time limit of 1 minute. The farther the participant moved, the higher the score. This Test Game enabled us to objectively measure the users' understanding of the FER system, through the score.

\subsection{Study Design}

The study was designed as a between-subject study with 2 cohorts:

\begin{enumerate}
    \item \textbf{AutoHint} Participants were provided with hints only (no explanations). Participants were instructed prior to the experiment that video data will be used to detect emotions which, in turn, trigger the hint system. No further details were provided. 
    \item \textbf{XAutoHint} Participants are exposed to global (mandatory) as well as local (optional) explanations of the FER system explained in Sec.~\ref{explanation-mechanism}.
\end{enumerate}

\subsection{Experimental Procedure}

Participants performed the experiment individually on a laptop in a private, well-lit room. An onboard camera was used to capture the video stream for processing. Participants used the onboard keyboard and a mouse to interface with the system. After a few practice runs, both cohorts proceeded to play the same set of maps for the \textbf{Explorer Game} with a randomized map order to control for any ordering effects. Participants in the \textbf{XAutoHint} cohort were able to view local explanations as described in the previous section. After playing the Explorer Game, participants  proceeded to play the \textbf{Test Game} which aims to evaluate their understanding of the FER. During the games, the experimenter exited the testing room to ensure that participants could express themselves freely and to reduce experimenter bias. After both games were completed participants answered the the Trust Scale~\cite{hoffman2018metrics}.

After the questionnaires were completed and prior to conclusion, participant exit interviews were conducted. Participants were asked to explain how the hint mechanism worked and which facial muscles or emotions were needed to trigger the hint mechanism in the exit interview.

Thematic analysis~\cite{braun2006using} was used to evaluate the interview responses and identify themes and patterns in the data. Following the thematic analysis method outlined by~\cite{braun2006using}, the qualitative data from the interviews were analysed in depth to extract common themes. We identified themes using a data-driven/inductive approach. The themes were updated throughout the review of all interview data. We reported the identified themes and added relevant extracts from participants' own `verbatim' transcripts within an analytic narrative \cite{adams2008qualititative}. This narrative of the qualitative analysis is presented in Sec. ~\ref{subsec:FER_understanding}. 

\subsection{Participants}
 
The study was conducted with a total of 20 participants, 10 participants in each cohort (5 M, 5 F). No significant differences were found between the cohorts in terms of age ($U = 44.5$, $p = 0.70$), gender ($U = 50$, $p = 0.97$), self-reported navigation capabilities ($U = 38$, $p = 0.38$) and self-reported emotional expressiveness/suppressiveness ($U = 42.5$, $p = 0.60$) using a Wilcoxon rank-sum test.

\section{Results}

\begin{figure}[tb]%
\centering
{{\includegraphics[width=0.7\linewidth]
{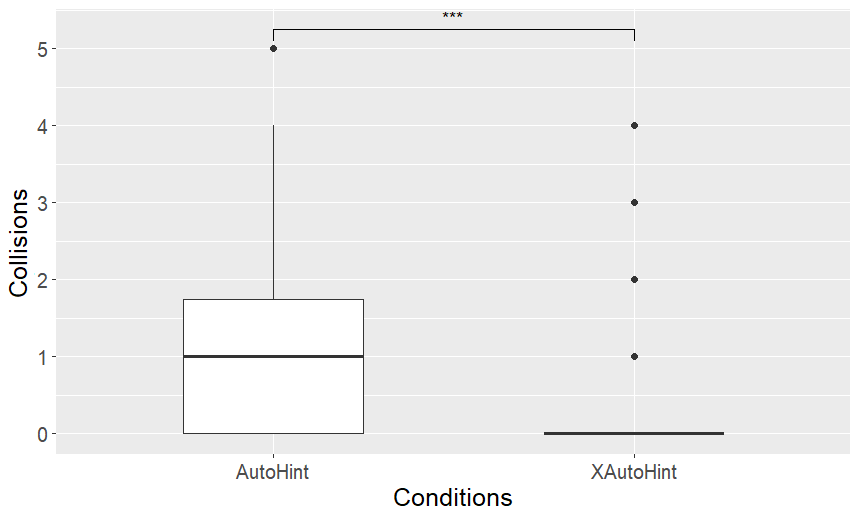} }}%
\caption{Boxplot on collisions during Explorer Game. Significantly less collisions for XAutoHint group compared to AutoHint group}%
\label{study2-collisions-fig}
\end{figure}

\subsection{Task Performance}

The in-game performance data for the Explorer Game was analysed to evaluate \textbf{H1} (\textit{A user will have better task performance when provided with explanations}). Since the hints helped users avoid collision with enemy agents we compared the amount of collisions experienced by participants in each cohort. We found a statistically significant difference in the number of collisions ($W = 5434.5$, $p < 0.001$) between both cohorts with the XAutoHint cohort experiencing a significantly lower number of collisions (average collision of 0.32) compared to the AutoHint cohort (average collision of 0.99) as can be seen in Fig.~\ref{study2-collisions-fig} \footnote{This significance held during post-analysis when  outliers were removed ($W = 3967.5$, $p-value < 0.001$).}). The size of the effect was medium with a Cohen's d value of 0.681.

When analysing overall game performance as determined by the game scores (amount of fuel leftover at the end of the game) and completion rate we found no significant difference for the scores ($W = 3522$, $p = 0.13$) and completion rate ($W = 53$, $p = 0.84$) using the Wilcoxon rank-sum test. This, however, is not necessarily a reflection on the helpfulness of the hints since fewer collisions does not directly correlate to less fuel usage. The hints allowed participants to avoid collisions (which have a higher penalty), however the extra moves made to avoid collisions still costs fuel (lower penalty) and may have lowered the scores of participants in the \textbf{XAutoHint} group which resulted in no significance being detected for the overall scores or completion rate.

\subsection{FER System Understanding}\label{subsec:FER_understanding}

\begin{figure}[tb]%
\centering
\subfloat[\label{study2-hinttriggers-fig}Significantly fewer hints were triggered in the Explorer Game for the AutoHint cohort compared with the XAutoHint cohort.]{{\includegraphics[width=0.45\linewidth]
{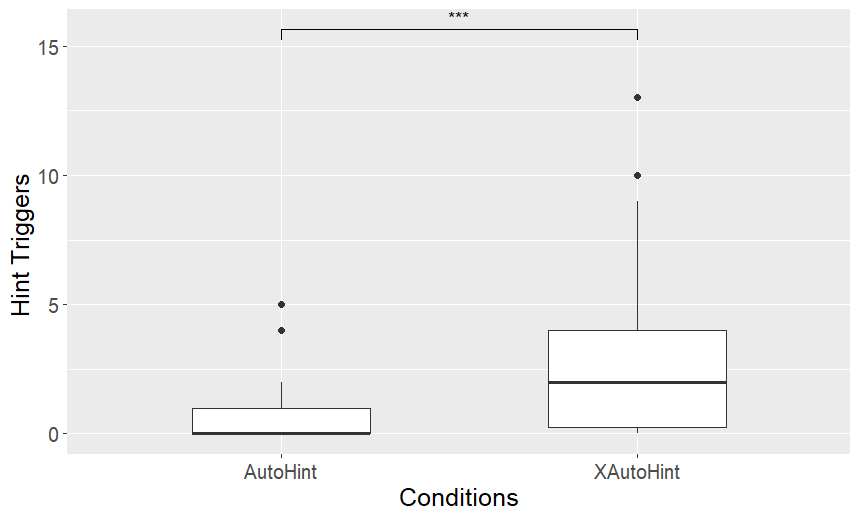} }}%
~\quad
\subfloat[\label{study2-testappscore-fig}Median Test Game Score for the AutoHint cohort is significantly lower than the XAutoHint cohort.]{{\includegraphics[width=0.45\linewidth]
{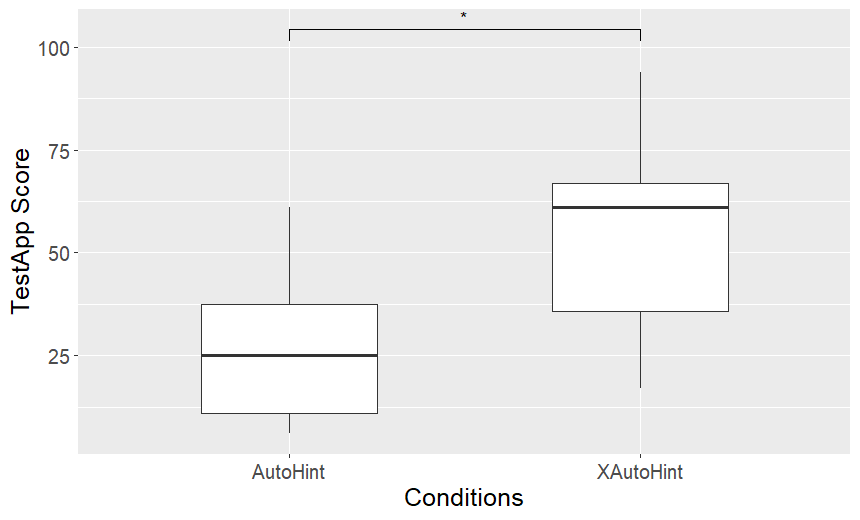} }}%
\caption{Boxplots on hint triggers and Test Game score.}%
\label{HintTriggersAndTestAppScore}%
\end{figure}

We next attempt to address \textbf{H2} (\textit{A user will report and demonstrate better understanding of the system when provided with explanations}). This can be ascertained by comparing the number of hints triggered during game execution between both cohorts. A higher number of triggered and accepted hints indicates a better understanding of which expressions can trigger the system and which cannot. The number of hints triggered in the Explorer Game is shown to be significantly higher in the XAutoHint cohort ($W = 1906.5$, $p < 0.001$) as can be seen in Fig.~\ref{study2-hinttriggers-fig} \footnote{This significance held during post-analysis when  outliers were removed ($W = 1636.5$, $p-value < 0.001$).}. The average number of hint triggers for the XAutoHint cohort was 2.76 whereas for the AutoHint cohort it was 0.66. The size of the effect was large with a Cohen's d value of 1.03. We further evaluated the overall scores of the Test Game which is determined by the number of times the hint system was activated. When comparing between the AutoHint and XAutoHint cohorts, the scores for the Test Game are significantly higher for participants in the XAutoHint cohort ($W = 17.5$, $p = 0.015$) indicating more hints were triggered, as seen in Fig.~\ref{study2-testappscore-fig}. The average score was 53.8 for the XAutoHint cohort and 26.7 for the AutoHint cohort. The size of the effect was large with a Cohen's d value of 1.23.

Qualitatively, we compare the interview responses from participants explaining how the hint function works in their own words. For the \textbf{AutoHint} group, the themes found indicated that participants in this group have a `vague' and sometimes `incorrect understanding' of what emotions/expressions seem to trigger the hint mechanism. Although 4 out of 10 participants in this group mentioned that the ``\textit{eyes getting smaller, not happy face, and squeeze eyebrows}'' (P1, P5, P13) seem to trigger the hint, 9 out of 10 of the participants also mentioned a mix of different thoughts that indicate incorrect understanding of how the hints work. For example, some mentioned that a ``\textit{thinking, anxious, concerned or even confused face}'' (P3, P9, P13, P15, P17, P19) seemed to work in triggering the hints. For the participants in the \textbf{XAutoHint} group however, the understanding of the system is more `concise' and `accurate`. All 10 participants mentioned that the emotions that worked for them were  ``\textit{mainly negative emotions}'' (P6) such as ``\textit{disgust, surprise, anger, fear}'' (P8, P10, P12, P14) whilst others went even further to mention facial movements that worked for them such as ``\textit{raising/lowering eyebrows, tightening their mouth or opening it, and opening their eyes wider}'' (all \textbf{XAutoHint} participants).

\subsection{User Trust}

To evaluate \textbf{H3} (\textit{A user will report higher trust towards the system when provided with explanations}), we analysed perceived trust as obtained from the survey data of the Trust Scale~\cite{hoffman2018metrics}. The overall Trust Scale scores are significantly higher for the XAutoHint cohort ($W = 17$, $p = 0.014$). The size of the effect was large with a Cohen's d value of 1.29. 

A further indication of demonstrated trust towards the FER is obtained by analysing the number of hints ``accepted''. To compare the hint acceptance (measured by Yes/No button clicks) we performed a Welch's T-test, as the number of hints triggered were different for both groups and equal sample tests could not be carried out. The results indicate that participants demonstrated a significantly higher hint ``acceptance'' ($t = -5.7162$, $df = 40.694$, $p < 0.001$) in the \textbf{XAutoHint} condition \footnote{This significance held during post-analysis when  outliers were removed ($t = -7.7042$, $df = 32$, $p < 0.001$).}. The analysis of the data supports H3, indicating that users have a higher trust towards the system when provided with explanations. We also note that participants in the \textbf{XAutoHint} condition had higher hint ``compliance'' ($t = -2.3647$, $df = 19.77$, $p = 0.028$) \footnote{This significance held during post-analysis when  outliers were removed ($t = -3.2929$, $df = 18$, $p = 0.004$).}.

\section{Discussion}\label{subsec:discussion}

In this study we sought to understand how explanations of a Facial Expression Recognition (FER) used to provide assistance affect user perception, reliance and performance. We further investigated how the explanations improved user understanding and subsequent use of the facial expression triggered assistance system.

\textbf{\textit{Explanation effect on FER model understanding and user control: }} To evaluate the influence of the explanations on user understanding (\textbf{H2})  and control (\textbf{H1}) of the FER model, we examined the number of times the hint function was triggered between both cohorts. We found that  participants in the \textbf{XAutoHint} group had a significantly higher hint trigger rate when compared to participants in the \textbf{AutoHint} group, as seen in Fig. \ref{study2-hinttriggers-fig}. The higher hint trigger rate as well as hint ``acceptance'' shows that participants in the \textbf{AutoHint} group triggered the hints intentionally to aid their next move and avoid collisions (Fig.~\ref{study2-collisions-fig}). The Test Game scores for participants in the \textbf{XAutoHint} group were also significantly higher when compared to participants in the \textbf{AutoHint} group (Fig.~\ref{study2-testappscore-fig}). These results  empirically show that when using an imperfect,  static FER model, explanations improve user understanding and subsequent control of the model. As static models do not change, explanations can provide insights to the user on how to better utilize these models for their benefit and improve their usability in real world scenarios.

By further analysing the local explainer data, we found that when the local explainer was clicked, the hints ``accepted'' were much higher for the \textbf{XAutoHint} group  as compared to the \textbf{AutoHint} group. In terms of the hint ``compliance'', when the local explainer was clicked, there was no significant difference in the rates between the two groups. This  indicates that local explanations  had an effect on user trust towards the \textit{accuracy} of the hint triggering mechanism, however did not effect user trust towards the \textit{quality} of the move recommendations. This is a particularly interesting finding as it illustrates the possible different effects that global and local explanations have towards a user's understanding, control and trust of a FER model and the HCI system adopting it, and thus should be further investigated.

\textbf{\textit{Explanation effect on user trust: }} When considering explanation influence on user trust (\textbf{H3}), we analyse the perceived trust through subjective surveys as well as demonstrated trust through objective in-game metrics. The purpose of analysing both demonstrated trust and perceived trust is to determine if the practical action taken by the user is the same as the user's perceived trust, which may not always align\mbox{~\cite{miller2022we}}. The results of both these analyses show that the \textbf{XAutoHint} group exhibited a higher degree of trust compared to the \textbf{AutoHint} group. This supports the claim that explanations can engender greater trust in a system through understanding  how the model functions. Introducing explanations is useful to promoting the user adoption of FER-based applications to their advantage. However, care must be taken to reduce over or under reliance or trust \mbox{\cite{miller2022we}}. As the data captured on the hint compliance shows, the \textbf{XAutoHint} group complied with the recommended next move to be taken more often. This could be attributed to the fact that hints were  provided in a timely manner when required. Therefore, participants were more likely to follow the recommended move in order to avoid collisions. For the participants in the \textbf{AutoHint} cohort, the hints were not provided at the right time due to a lack of user understanding as to how to trigger hints through the relevant FAUs. We hypothesize that this is the reason participants were less likely to follow the recommended move. This results in user trust in the overall system to reduce. Future research should be carried out to ensure that the explanation mechanism engenders an appropriate level of trust in the FER. 

\section{Limitations and Future Work}

Our study should be viewed in light of the following limitations. The sample size for the study (N=20) was relatively small and therefore certain effects may not be apparent. However,  we still noted significant differences for the number of collisions, the number of hint triggers and the perceived and demonstrated trust between the two cohorts. In future, we intend to conduct validation studies with increased  sample size while maintaining an appropriate balance of demographics between different cohorts. DL methods have achieved improved performance in FER compared to the SVM model adopted in this work. Using FAUs as a form of explanation can be combined with these models. It is worth looking into replacing the SVM component to allow the interpretability method to be scaled to DL methods. And finally, we note that having users interact concurrently with the Local Explainer and with the task might affect user behaviour, potentially biasing participants' decisions. While necessary measures were taken to minimise this influence, such as pausing the timer while participants interacted with the Local Explainer and showing explanations on-demand, in future we plan to implement a similar review interface for both cohorts to further reduce its influence including basic game information without displaying FER explanations.

\section{Conclusion}

We proposed a novel explanation system for facial expressions recognition (FER) and investigated how explanations affect a user's task performance, system understanding, and trust when using a human-computer interaction (HCI) system that utilises FER to automatically trigger helpful hints. Through an improved understanding (\textbf{H2}) of the system, participants in our user study showed better task performance (\textbf{H1}) in two screen-based navigation games. This indicates that users provided with explanations have better control of the system compared to those who did not receive explanations. The survey and in-game data also suggests that users provided with explanations have a higher degree of perceived and demonstrated trust towards the system (\textbf{H3}). Our results indicate that explanations on a FER system help to engender a greater understanding and trust by the users towards an emotion-aware HCI system.

\section*{Ethical Impact Statement}

The experimental protocols were reviewed and approved by the Monash University Human Ethics Review Committee before participant recruitment and user studies (project ID 37086). Full participant consent was obtained and all data in non-identifiable. The majority of the participants were recruited from our University therefore may be biased towards a younger age with  higher education background. We did not collect cultural background of participants.

\bibliographystyle{plain}

{\footnotesize
\bibliography{bibliography}}

\end{document}